\documentclass{aa} 

\usepackage{graphicx}
\usepackage{txfonts}
\usepackage{natbib}
\usepackage{capt-of}
\begin{document} 
\renewcommand{\thefootnote}{\alph{footnote}}

   \title{Search for methylamine in high mass hot cores}

   \subtitle{}

   \author{N.F.W. Ligterink
          \inst{1,2}, E.D. Tenenbaum \inst{1}
          \and
          E.F. van Dishoeck\inst{1,3}
          }

   \institute{Leiden Observatory, Leiden University, PO Box 9513, 2300 RA Leiden, The Netherlands\\
              \email{ligterink@strw.leidenuniv.nl}
         \and
             Raymond and Beverly Sackler Laboratory for Astrophysics, Leiden Observatory, Leiden University, PO Box 9513, 2300 RA Leiden, The Netherlands\         		\and
             Max-Planck Institut f\"ur Extraterrestrische Physik (MPE), Giessenbackstr. 1, 85748 Garching, Germany\\             
             }

   \date{Received August 21, 2014}

 
  \abstract
   {}
{We aim to detect methylamine,
   CH$_{3}$NH$_{2}$, in a variety of hot cores and use it as a test for
   the importance of photon-induced chemistry in ice mantles and
   mobility of radicals. Specifically, CH$_3$NH$_2$ cannot be formed
   from atom addition to CO whereas other NH$_2$-containing molecules
   such as formamide, NH$_2$CHO, can.}
{Submillimeter spectra of several massive hot core
   regions were taken with the James Clerk Maxwell Telescope.
Abundances are determined with the rotational diagram
   method where possible.}
{Methylamine is not detected,
giving upper limit column densities between 1.9 $-$ 6.4 $\times$
10$^{16}$ cm$^{-2}$ for source sizes corresponding to the 100 K envelope 
radius. 
Combined with previously obtained JCMT data analyzed 
in the same way, abundance ratios of CH$_{3}$NH$_{2}$,
NH$_{2}$CHO and CH$_{3}$CN with respect to each other and to CH$_{3}$OH are
determined. These ratios are compared with Sagittarius B2 observations, where all species are detected, 
and to hot core models.}
{The
observed ratios suggest that both methylamine and formamide are
overproduced by up to an order of magnitude in hot core
models. Acetonitrile is however underproduced. The proposed chemical
schemes leading to these molecules are discussed and 
reactions that need further laboratory
studies are identified.
The upper limits obtained in this paper can be used to guide future
  observations, especially with ALMA.}


   \maketitle
%

\section{Introduction}

Complex organic molecules are thought to be formed primarily on dust
grains in dense cores, see reviews by
\citet{dishoeckherbst2009} and \citet{caselliceccarelli2012}. Before the onset of star formation, the
atomic and molecular reservoir is contained in large dark clouds. Due to
the high densities ($\geq$ 10$^{4}$ cm$^{-3}$) and low temperatures
(10 K) reached in these environments, gas-phase species will freeze
out on sub-micron sized grains forming ice mantles on timescales
shorter than the lifetime of the cloud. It is here that atoms and
molecules can potentially react with each other to form the zeroth
order ice species like ammonia, methane, water and methanol. UV
radiation interacts with these ice mantles by dissociating molecules to produce radicals and by photodesorbing species
back to the gas phase. If these radicals are sufficiently mobile, they can find
each other on the grain and react to form even more complex first
generation (organic) species \citep{garrodherbst2006}. However, it is not
entirely clear if UV radiation is essential to form these complex
molecules or whether they can also be formed just by thermal
processing and atom bombardment of solid CO with C, N and O atoms
\citep{tielenscharnley1997}.

In this context methylamine, CH$_{3}$NH$_{2}$, is a particularly interesting molecule, since its formation is hypothesised by \citet{garrod2008} to be completely dependent on radicals produced by UV photons, and is one of the few molecules that can definitely not be produced in the routes starting from solid CO:
\\ \\
CH$_{4} + h\nu \rightarrow $CH$_{3}^{\bullet} + $H
\\ \\
NH$_{3} + h\nu \rightarrow $NH$_{2}^{\bullet} + $H
\\ \\
CH$_{3}^{\bullet} + $NH$_{2}^{\bullet} \rightarrow $CH$_{3}$NH$_{2}$
\\ \\
These radicals can form in the ice mantles in the dark cloud or in the
protostellar phase through cosmic-ray induced photons and/or UV
photons from the protostar. After gravitational collapse of the cloud
and formation of a protostar, the dust around it will start to warm
up. The increased temperature will cause the radicals to become mobile
on the grains and react with each other, forming methylamine. Further
heating will evaporate the formed methylamine from the grain and
raise its gas-phase abundance.  

Another interesting amine-containing molecule is formamide,
NH$_{2}$CHO. This is so far the most abundantly observed
amine-containing molecule \citep[e.g.,]{halfen2011,bisschop2007},
making it an interesting molecule to compare with other amines like
methylamine. In contrast with CH$_3$NH$_2$, this molecule can possibly be
produced by reactions of H and N with solid CO. The comparison of the
abundances of these two species could potentially give more
information about the relative importance of UV-induced versus thermal
grain surface reactions.

Hot cores are particularly well-suited to study methylamine. These
high mass star-forming regions reach high temperatures between 100 to
300 K and are known for their rich complex organic chemistry
\citep{walmsley1992,dishoeckblake1998, tielenscharnley1997,
  ehrenfreund2000,caselliceccarelli2012}. The ice covered grains move inwards to
the protostar and will heat up. When sufficient temperatures are
reached, molecules will start to desorb depending on their respective
binding energies. Less abundant molecules mixed with water ice will
desorb together with water around 100 K.

Previous detections of methylamine have all been made toward the
galactic center. \citet{kaifu1974} first detected CH$_{3}$NH$_{2}$ in
Sagittarius B2 and Orion A. Later that same year \citet{fourikis1974}
reported the detection of methylamine in the same sources, but with a
different telescope. Much more sensitive surveys by
\citet{turner1991}, \citet{nummelin2000}, \citet{halfen2013},
\citet{belloche2013} and \citet{neill2014} also all detected
methylamine lines toward SgrB2, with typical inferred abundance ratios
with respect to NH$_{2}$CHO between 0.5 to 3. No detections of
methylamine have been reported in sensitive surveys with modern
detectors toward Orion, however
\citep{blake1987,turner1991,sutton1995,schilke1997,crockett2014}.

To study the importance of UV processing of ice-covered dust grains,
we present the results of searches for methylamine in a
number of hot cores (see Table \ref{sourcelist}). These results are
combined and compared with data from \citet{bisschop2007} and
\citet{isokoski2013}, which were taken toward the same hot cores with
the same telescope and analysis method and include detections of
NH$_{2}$CHO and other nitrogen-containing species. In Section
\ref{obs} the observational details are given, followed by the
analysis method in Section \ref{data}. Section \ref{result} summarizes all
the results of our analysis and these are discussed in Section
\ref{discus}. Finally conclusions are drawn in Section \ref{con}.


\section{Observations}
\label{obs}

   \begin{table*}
      \caption[]{Source list and source parameters}
         \label{sourcelist}
     $$ 
         \begin{tabular}{l r r r r r r r r r r}
            \hline
            \noalign{\smallskip}
            Source & RA	& Dec & $\theta_{\rm{S}}^{a}$ & $\theta_{\rm{B}}^{a}$ & $L^{a}$ & $d^{a}$ & $V_{\rm{LSR}}$ & $\Delta V$ & $\delta \nu$ & $RMS$  \\
            & J2000 & J2000 & AU & AU & $L_{\odot}$ & (kpc) & (km s$^{-1}$) & (km s$^{-1}$) & (km s$^{-1}$) & (mK)  \\
            \noalign{\smallskip}
            \hline
            \noalign{\smallskip}
                  
         	AFGL 2591	& 20:29:24.60 & $+$40:11:18.9	& 1800  & 21000 & 2.0E+04 & 1.0 & -5.5 & 4.0 & 1.28 & 10	\\
            G24.78		& 18:36:12.60 & $-$07:12:11.0	& 13000 & 162000 & 7.9E+05 & 7.7 & 111.0 & 6.3	& 1.28 & 9 \\
           	G31.41+0.31	& 18:47:34.33 & $-$01:12:46.5	& 7840 & 166000 & 2.6E+05 & 7.9 & 98.7 & 7.3 & 1.28 & 7	\\
           	G75.78		& 20:21:44.10 & +37:26:40.0 & 5600 & 86100 & 1.9E+05 & 4.1 & -0.04 & 5.6 & 1.28 & 9	\\
           	IRAS 18089-1732	& 18:11:51.40 & $-$17:31:28.5	& 2750 & 49000 & 3.2E+04 & 2.3 & 33.8 & 4.5 & 1.28 & 9	\\
           	IRAS 20216+4104	& 20:14:26.40 & +41:13:32.5	& 1753 & 34400 & 1.3E+04 & 1.6 & -3.8 & 6.0 & 1.28 & 10	\\
           	NGC 7538 IRS1	& 23:13:45.40 & +61:28:12.0	& 4900 & 58800 & 1.3E+05 & 2.8 & -57.4 & 4.0 & 1.28 & 10	\\
           	W3(H$_{2}$O)	& 02:27:04.60 & +61:52:26.0	& 2400 & 42000 & 2.0E+04 & 2.0 & -46.4 & 5.0 & 1.28 & 11 \\
           	W 33A		& 18:14:38.90 & $-$17:52:04.0 	& 4500 & 84000 & 1.0E+05 & 4.0 & 37.5 & 4.9 & 1.28 & 11	\\
           
            \noalign{\smallskip}
            \hline            
         \end{tabular}
     $$ 
     $^{a}$ Data taken from \cite{bisschop2007} and \cite{isokoski2013}.\\ 
   \end{table*}

Observations were performed with the James Clerk Maxwell Telescope
(JCMT) \footnote{The James Clerk Maxwell Telescope is operated by the
  Joint Astronomy Centre on behalf of the Science and Technology
  Facilities Council of the United Kingdom, the National Research
  Council of Canada, and (until 31 March 2013) the Netherlands
  Organisation for Scientific Research.}  on the sources listed in
Table \ref{sourcelist} between July 2010 and August 2011. The sources
were selected based on their particularly rich chemistry, being
isolated, having narrow line widths to prevent line confusion and on
their relatively nearby distance \citep{bisschop2007, fontani2007,
  rathborne2008, isokoski2013}.

\cite{nummelin1998} detected methylamine emission lines between 218 to
263 GHz toward Sgr B2N. Therefore the RxA3 front-end double side band
receiver, functioning between 210 to 276 GHz, was chosen to observe
the hot cores. The 250 and 1000 MHz wide back-end ACSIS configurations
were used.  A number of methylamine transitions covering a range of
excitation energies were selected in this frequency range based on
high Einstein $A$ coefficients and lack of line confusion
(Table \ref{methylamine}). However, not all transitions were observed
for all sources. The 235735 MHz transition was only recorded for
W3(H$_{2}$O) and the 260293 MHz transition only toward W3(H$_{2}$O)
and NGC 7538 IRS1.

Because double side band spectra were obtained, our spectra contain
transitions from two different frequency regimes superposed. To
disentangle lines from the two side bands, each source was observed
twice with an 8 MHz shift in the local oscillator setting between the
two observations. 
This allows each transition to be uniquely assigned to either of the
two side bands.

In the 230 GHz band, the JCMT has a beam size ($\theta_{B}$) of
20-21$''$. Spectra were scaled from the antenna temperature scale,
$T_{\rm{A}}^{*}$, to main beam temperature, $T_{\rm{MB}}$, by using
the main beam efficiency of 0.69 at 230 GHz. Integration times were
such that $T_{\rm{RMS}}$ is generally better than 10 mK for data
binned to 1.3 km s$^{-1}$ velocity bins. Noise levels were improved by
adding the shifted spectra together in a narrow frequency region around the CH$_{3}$NH$_{2}$ lines, effectively doubling the
integration time.

   \begin{table}
      \caption[]{Methylamine transitions observed in this study$^{a}$}
         \label{methylamine}
     $$ 
         \begin{tabular}{c c r c r}
            \hline
            \noalign{\smallskip}
            Transition &  Freq	& $E_{\rm{up}}$	& $A$ & $g_{\rm{up}}$  \\
            & (MHz)	& (K)	& (s$^{-1}$) &		\\
            \noalign{\smallskip}
            \hline
            \noalign{\smallskip}
            
            4$_{2}$ $\rightarrow$ 4$_{1}$$^{b}$ &  229310.298 & 36.9 & 1.32E-05 & 108     \\
            7$_{2}$ $\rightarrow$ 7$_{1}$$^{b}$ &  229452.603 & 75.5 & 5.88E-06 & 60    \\
            8$_{2}$ $\rightarrow$ 8$_{1}$$^{c}$ &   235735.037 & 92.8 & 6.13E-05 & 204    \\
            6$_{2}$ $\rightarrow$ 6$_{1}$$^{b}$ &  236408.788 & 60.8 & 5.94E-05 & 52    \\
            2$_{2}$ $\rightarrow$ 2$_{1}$$^{b}$ &  237143.530    & 22.0 & 3.82E-05 & 60 \\
            10$_{2}$ $\rightarrow$ 10$_{1}$$^{d}$ &  260293.536 & 132.7 & 2.26E-05 & 52    \\
            \noalign{\smallskip}
            \hline
         \end{tabular}
     $$ 
     $^{a}$ Data from JPL database for molecular spectroscopy.\\
     $^{b}$ Transition observed in all sources.\\
     $^{c}$ Only observed in W3(H$_{2}$O).\\
     $^{d}$ Only observed in W3(H$_{2}$O) and NGC 7538 IRS1.\\
     
   \end{table}
   
\section{Data analysis}
\label{data}

   \begin{figure*}
    \includegraphics[width=\hsize]{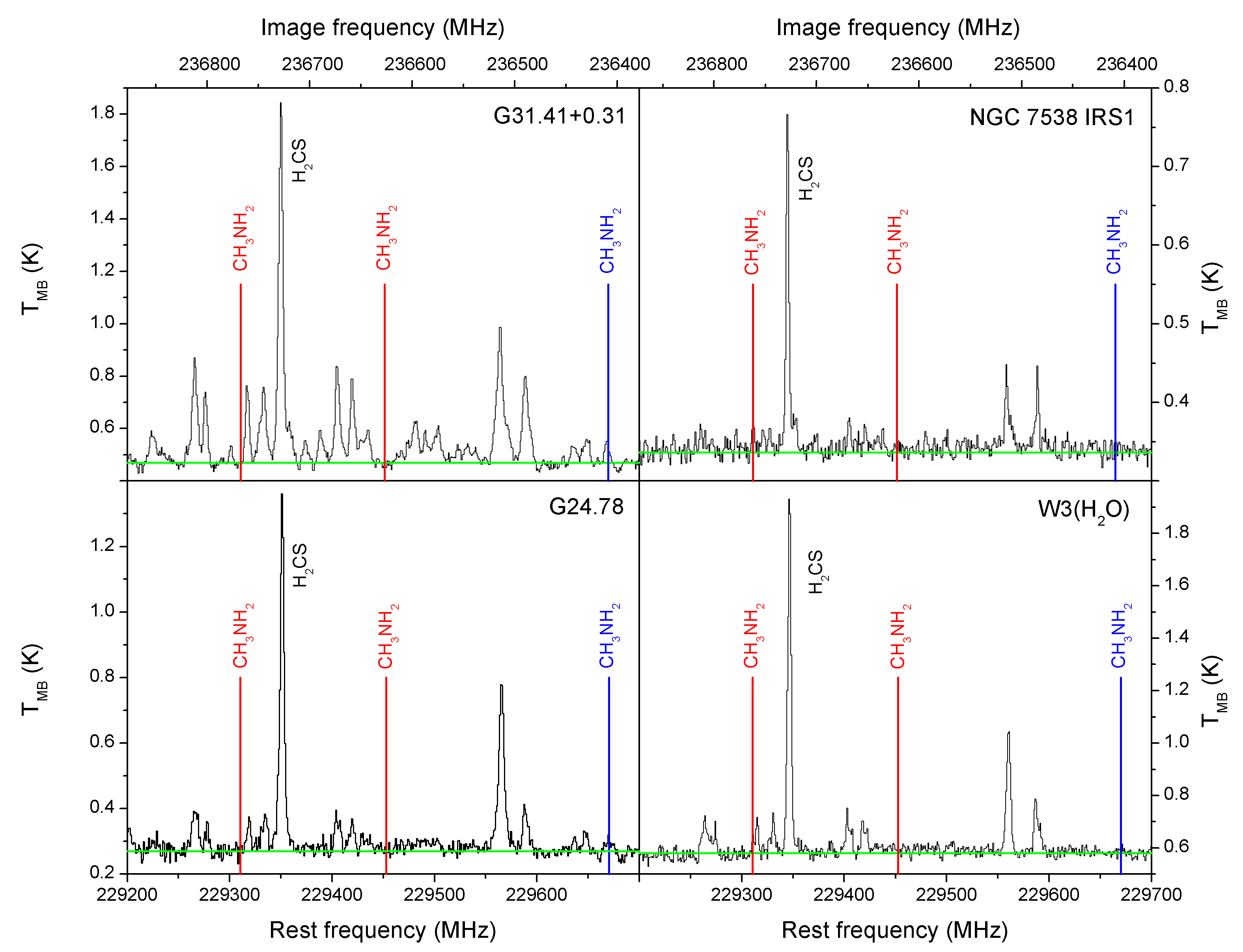}
      \caption{JCMT spectra of the massive hot cores G31.41+0.31,
        G24.75, NGC 7538 IRS1 and W3(H$_{2}$O). The 229310 and 229452
        MHz transitions in the lower sideband are indicated in red and
        that at 236408 MHz in the upper sideband in blue. In green is
        the baseline, obtained by fitting line free portions of the spectrum. In all spectra the H$_{2}$CS 7$_{1}$
        $\rightarrow$ 6$_{1}$ transition at 236726 MHz is fitted
        to determine the typical linewidth in the sources,
        as listed in Table \ref{sourcelist}.  }
         \label{spechc}
   \end{figure*}

To analyse the data, exactly the same method as described by
\cite{bisschop2007} and \cite{isokoski2013} was used. It will be
shortly reiterated here.  The hot core spectra corrected for source
velocity were analysed with the "Weeds" extension \citep{maret2011} of
the Continuum and Line Analysis Single-dish Software (CLASS\footnote{\textbf{http://www.iram.fr/IRAMFR/GILDAS}}) coupled
with the Jet Propulsion Laboratory (JPL\footnote{\textbf{http://spec.jpl.nasa.gov}}) database for molecular
spectroscopy \citep{pickett1998}. Focus was on identifying the transitions of methylamine
listed in Table \ref{methylamine}, but other lines in the spectra were
measured as well (see Table \ref{fulltrans} Appendix). After each
positive identification the integrated main-beam temperature,
\textit{$\int T_{ \rm{MB}}dV$}, was determined by gaussian fitting of
the line. From the integrated main-beam intensity the column density
\textit{$N_{\rm{up}}$} and thus the \textit{beam-}averaged total
column density \textit{$N_{\rm{T}}$} could be determined, assuming
Local Thermodynamic Equilibrium (LTE) at a single excitation
temperature $T_{\rm{rot}}$:\\
	\begin{eqnarray}
	\label{colden}
	\frac{3k\int T_{\rm{MB}}dV}{8\pi^{3}\nu\mu^{2}S} = \frac{N_{\rm{up}}}{g_{\rm{up}}} = \frac{N_{T}}{Q(T_{\rm{rot}})}e^{-E_{\rm{up}}/T_{\rm{rot}}} 
	\end{eqnarray} 
   \\
where \textit{$g_{\rm{up}}$} is the level degeneracy, $k$ the Boltzmann constant, \textit{$\nu$} the transition frequency, \textit{$\mu$} the dipole moment and \textit{S} the line strength. $Q(T_{\rm{rot}})$ is the rotational partition function and $E_{\rm{up}}$ is the upper state energy in Kelvin. 

In case of a non-detection, $3\sigma$ upper limits were determined from the Root Mean Square (RMS) of the base line of the spectra in combination with the velocity resolution \textit{$\delta \nu$} and line width \textit{$\Delta V$}:
	\\
   \begin{eqnarray}
   \label{sigma}
   \sigma = 1.2\sqrt{\delta \nu \Delta V}\cdot RMS
   \end{eqnarray}
	\\   
\textit{$\Delta V$} is estimated from other transitions (see Table
\ref{sourcelist}) in the spectra, for example from the nearby
H$_{2}$CS 7$_{1} \rightarrow$ 6$_{1}$ transition, and assumed to be
the same for all transitions in the spectral range. A telescope flux
calibration error of 20\% is taken into account in the 1.2 factor. The
$3\sigma$ value is then used in the same way as the main-beam
intensity of detected lines to obtain the upper limit on the total
column density through Eq.~\ref{colden}.

Since no rotational temperature can be determined for a non-detection,
this has to be estimated. In the models of \citet{garrod2008} the peak
abundance temperatures for methylamine range from 117 to 124 K
depending on the model used. \citet{oberg2009} determined that
methylamine forms in CH$_{4}$/NH$_{3}$ UV irradiation experiments and
sublimates at 120 K. There is a small difference between laboratory
and hot core desorption temperatures, because of the pressure
difference between the two. Also, if CH$_{3}$NH$_{2}$ is embedded in
water ice the desorption temperature will probably be limited to
roughly 100 K, when water desorbs in space. Therefore
$T_{\rm{rot}}$ is assumed to be 120 K when methylamine lines could not
be identified, but the effects of lower and higher rotation
temperatures are explored as well.  

Correction for beam dilution is done in the same way as \citet{bisschop2007}:
   \begin{eqnarray}
   \label{eta}
   \eta_{BF} = \frac{\theta_{S}^{2}}{\theta_{S}^{2}+\theta_{B}^{2}} 
   \end{eqnarray}

resulting in the \textit{source-}averaged column density:

   \begin{eqnarray}
   \label{ns}
   N_{S} = \frac{N_{T}}{\eta_{BF}} 
   \end{eqnarray}

The beam diameter $\theta_{B}$ is set at 21". For the source diameter, $\theta_{S}$, values have been taken from \citet{bisschop2007} and \citet{isokoski2013} and constitute the area where the temperature is 100 K or higher and hot gas-phase molecules are present. Both beam and source diameters are listed in AU in Table \ref{sourcelist}. Using the CASSIS line analysis software \footnote{CASSIS has been developed by IRAP-UPS/CNRS (http://cassis.irap.omp.eu).} it was verified that the source-averaged column densities are still small enough that the observed lines are optically thin.

\section{Results and comparison with astrochemical models}
\label{result}
\subsection{CH$_3$NH$_2$ limits}

Figure \ref{spechc} presents examples of spectra obtained for our
sources, whereas Figure~\ref{2-2} in the Appendix shows the $2_2-2_1$
line in all sources. In general, no transitions of CH$_{3}$NH$_{2}$
are detected. Only one possible methylamine transition is identified
in G31.41+0.31 coincident with the 6$_{2}$ $\rightarrow$ 6$_{1}$ line at
236408 MHz, with an integrated intensity of 0.44 K
kms$^{-1}$. Following the procedure summarized in Section \ref{data},
a column density of 3.4 $\times$ 10$^{17}$ cm$^{-2}$ is inferred from
this line assuming $T_{\rm rot}=120$~K. However, modelling of the
spectrum shows that the other targeted CH$_3$NH$_2$ lines, 4$_{2}$
$\rightarrow$ 4$_{1}$ and 2$_{2}$ $\rightarrow$ 2$_{1}$, should have
comparable or even higher intensities if this identification is
correct (Figure \ref{tmb}). The 8$_{2}$ $\rightarrow$ 8$_{1}$ line should
be readily detected but was not observed toward G31.41+0.31.  This
makes it unlikely that the detected feature belongs to methylamine,
since we would expect to see at least two other CH$_3$NH$_2$
transitions in our spectrum.

\begin{figure}
   \centering
   \includegraphics[width=\hsize]{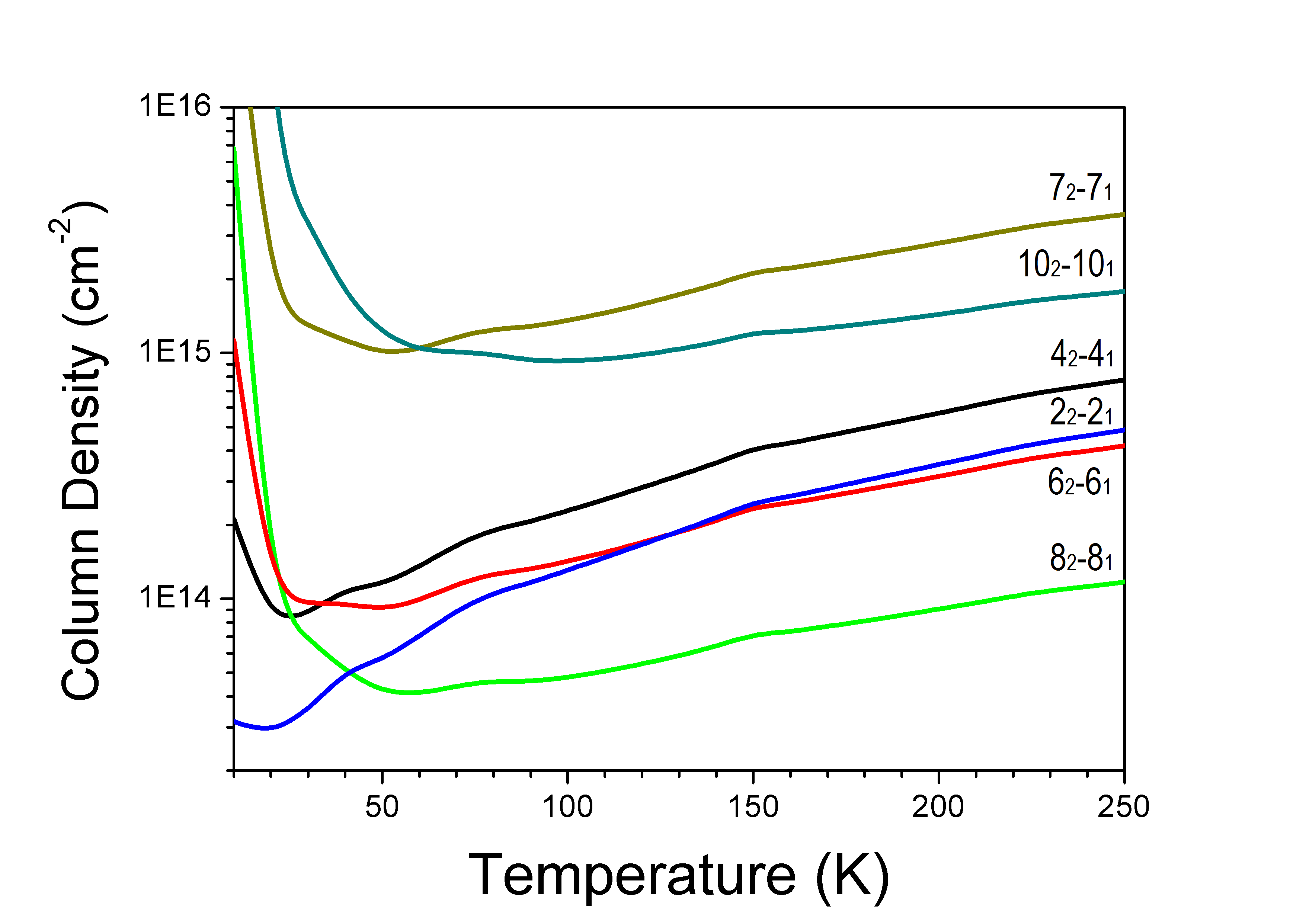}
      \caption{Column densities for the six methylamine transitions
        plotted versus temperature. This plot is made for a 3$\sigma$
        limit of 0.1 K kms$^{-1}$, as found for W3(H$_{2}$O). This
        figure demonstrates that the $8_2-8_1$ transition (green)
        gives the most sensitive limits on column density for the
        relevant range of excitation temperatures in hot cores, when
        observed. The other five transitions (4$_{2}$ $\rightarrow$
        4$_{1}$, black; 7$_{2}$ $\rightarrow$ 7$_{1}$, gold; 10$_{2}$
        $\rightarrow$ 10$_{1}$, cyan; 2$_{2}$ $\rightarrow$ 2$_{1}$,
        blue and 6$_{2}$ $\rightarrow$ 6$_{1}$, red) clearly imply
        higher column densities.  Only below 40 K does the 2$_{2}$
        $\rightarrow$ 2$_{1}$ line give lower column density limits.
      }
         \label{coldens}
   \end{figure}

In Figure \ref{coldens} upper limit column densities of the six investigated
transitions of methylamine are plotted versus rotational temperature
taking a typical 3$\sigma$ = 0.100 K kms$^{-1}$. At 120 K the 8$_{2}$
$\rightarrow$ 8$_{1}$, 235735 MHz transition gives the lowest limits
on the column densities, see Figure \ref{coldens}. However, since this
particular transition was only included in the observations for one
source, the second most sensitive transition at 120 K, 2$_{2}$
$\rightarrow$ 2$_{1}$, will be used (see Figure \ref{2-2} in the
Appendix for a blow-up of this particular spectral region in all
investigated sources). All following molecular ratios are based on CH$_{3}$NH$_{2}$ column densities obtained from this line, assuming T$\rm _{rot}$ = 120 K. The corresponding upper limits are presented in Table \ref{ratiosmeth}

\subsection{Abundance ratio comparison}

Combined with NH$_{2}$CHO and CH$_{3}$OH column densities
  from \citet{bisschop2007} and \citet{isokoski2013} derived in the
same way, abundance ratios for methylamine and formamide with respect
to each other and to methanol are calculated. These ratios are listed
in Table \ref{ratiosmeth}. Methanol is chosen as a reference since it
is the most readily observed complex organic molecule. Its
disadvantage is that some of the transitions have high optical depth
and that a cold component may be present \citep{isokoski2013}, but
this is circumvented by only taking the warm methanol column density
derived from optically thin lines. Abundances relative to
  methanol rather than H$_2$ are preferred since the H$_2$ column
  depends on extrapolation of dust models to smaller scales than
  actually observed \citep{bisschop2007}. Another point that needs to be taken into account is that the models of \citet{garrod2008} do show a slight overproduction of CH$_{3}$OH, which could influence the comparison between the ratios. Overall, the abundance
ratios are estimated to be accurate to a factor of a few.

It should be noted that methylamine and formamide have
  significantly different dipole moments (1.31 and 3.73 Debye
  respectively) and could therefore be excited in different
  ways. Formamide has a larger critical density than methylamine, so
  the situation could arise where the critical density is not reached
  for formamide or even both molecules. The corresponding excitation
  temperatures will then be lower. In particular, the situation in
  which the critical density is not reached for formamide but is for
  methylamine, could affect the inferred ratios.  As can be seen from
  Figure \ref{coldens}, if $T_{\rm rot}$ drops from 120 to 50 K, the
  column density drops by a factor of a few, depending on
  transition. If $T_{\rm rot}$ were 120 K for methylamine but 50 K for
  formamide, the observed column density of formamide would be lower
  than that listed here and thus result in a higher
  CH$_{3}$NH$_{2}$/NH$_{2}$CHO ratio. We note, however, that there is
  no observational evidence that $T_{\rm rot}$ is systematically lower
  than 100 K for formamide \citep{bisschop2007}.

Table~\ref{ratiosmeth} includes the observational results toward Sgr
B2, the only source where methylamine is firmly detected, from
\citet{turner1991}, \citet{belloche2013} and \citet{neill2014}.  These
results, obtained over the course of more than two decades, agree well
with each other within the estimated uncertainties due to slightly
different adopted source sizes.  \citet{nummelin2000} also detect
methylamine in their Sgr B2 survey but find a surprisingly small beam
filling factor and consequently very large column density compared
with most other complex organic molecules. If their beam filling
factor for CH$_3$NH$_2$ is taken to be the same as for NH$_2$CHO, the
\citet{nummelin2000} ratios are more in line with those derived by
\citet{turner1991}, \citet{belloche2013} and \citet{neill2014}. The
non-detections of methylamine toward the chemically rich and well
studied Orion hot core imply abundance limits that are at least a
factor of 5 lower than for SgrB2 \citep{neill2014,crockett2014}.

Table~\ref{ratiosmeth} also contains the model results from
\cite{garrod2008}, who present three hot core models which differ from
each other by their warm-up timescale from 10 to 200 K. The timescales
for F(ast), M(edium) and S(low) are $5\times10^{4}$,
$2\times10^{5}$ and $1\times10^{6}$ years, respectively, and start after the cold collapse phase. In the slow models more time is spent in the warm-up
phase where radicals are mobile. Values used in this comparison are
taken from the so-called reduced ice composition, where cold phase
methane and methanol abundances were modified to match observations of
these ices toward W33A, NGC 7538 IRS9 and Sgr A*, see
\citet{gibb2000b}. Another comparison can be made with the gas-phase
abundances in protoplanetary disk models of \citet{walsh2014} which
have similar or higher densities and temperatures as in protostellar
cores. Their ratios range from 7.2 $\times$ 10$^{-1}$ to 6.5 $\times$
10$^{-2}$ for CH$_{3}$NH$_{2}$/CH$_{3}$OH, 4.2 $\times$ 10$^{-1}$ to
1.5 for CH$_{3}$NH$_{2}$/NH$_{2}$CHO and 1.7 to 8.8 $\times$ 10$^{-2}$
for the NH$_{2}$CHO/CH$_{3}$OH. These ratios are close to the
predicted values of \citet{garrod2008} listed in
Table~\ref{ratiosmeth}, which may be partly due to using the same
surface-chemistry network.

From Table \ref{ratiosmeth} several trends become apparent for our
results. The CH$_{3}$NH$_{2}$/NH$_{2}$CHO limits lie about an order of
magnitude above model values whereas the CH$_{3}$NH$_{2}$/CH$_{3}$OH
limit approximately matches with theoretical predictions.  Because the
observed values are actually 3$\sigma$ upper limits, this suggests
that models overproduce CH$_3$NH$_2$. For the sources with the most
stringent limits, such as G31.41+0.31 and the $8_2-8_1$ line in
W3(H$_{2}$O), the CH$_{3}$NH$_{2}$/CH$_{3}$OH limits are comparable or
even lower than the abundance ratios for Sgr B2. The third ratio,
NH$_{2}$CHO/CH$_{3}$OH, is also found to be lower than the models by
up to one order of magnitude.  

Close inspection of the \citet{bisschop2007} data shows that
  the NH$_2$CHO column densities may have larger uncertainties than
  quoted in their figures and tables. We have therefore re-analysed
  all NH$_2$CHO data from that paper taking larger uncertainties into
  account. In general, this leads to lower NH$_2$CHO column
  densities. Even using the upper limits from this
  re-analysis as well as those from \citet{isokoski2013} (which were
  obtained with generous error bars), the
  NH$_{2}$CHO$_{upper}$/CH$_{3}$OH ratios are significantly lower than
  the models. This suggests that
both the methylamine and formamide abundances are too high in the
models.

The Sgr B2 detections tend to have lower
  CH$_3$NH$_2$/NH$_2$CHO and CH$_3$NH$_2$/CH$_3$OH ratios than our
  upper limits and are also somewhat below the models, but generally
  do not differ more than a factor of a few. The SgrB2
  NH$_2$CHO/CH$_3$OH ratios are also closer to the models results, at
  least for the faster models. However, the Orion Compact Ridge
  NH$_2$CHO/CH$_3$OH value from \citet{crockett2014} is comparable to
  that found for our sources and clearly lower than the
  models. Further observations are needed to determine to what extent
  Sgr B2 is a special case.

To further elucidate the differences between theory and our and the
Sgr B2 observations, an additional analysis was made of the
CH$_{3}$NH$_{2}$/CH$_{3}$CN ratio. These results are listed in Table
\ref{ratioscya}. Acetonitrile is produced in the gas-phase, but more
abundantly on grains: an important route to its formation is via
CH$_{3}^{\bullet}$ + CN$^{\bullet} \rightarrow$ CH$_{3}$CN, according
to \citet{garrod2008}. This would mean that both molecules compete for
the methyl radical on the surface, thus relating the two molecules.

Our observed ratios involving CH$_3$CN are clearly at odds with the
theoretical predictions. The observed CH$_3$NH$_2$/CH$_3$CN ratios are
in most cases an order of magnitude lower than theory and approach the
observed ratios for Sgr B2. However, the observed
CH$_{3}$CN/CH$_{3}$OH ratios are 1-2 orders of magnitude higher than
theoretical predictions. Both these cases point to CH$_3$CN being
underproduced in the models.

Finally, abundance ratios, with some notable exceptions, do not vary more than an order of
magnitude between different sources, as also found by \citet{bisschop2007} for other species.

   \begin{table*}
      \caption[]{Upper limit column densities and abundance ratios for methylamine.}
         \label{ratiosmeth}
         $$
         \begin{tabular}{l l l l l l}
            \hline
            \noalign{\smallskip}
            Source & $N_{\rm S,CH_{3}NH_{2}}$  & CH$_{3}$NH$_{2}$/NH$_{2}$CHO & CH$_{3}$NH$_{2}$/CH$_{3}$OH &  NH$_{2}$CHO/CH$_{3}$OH & NH$_{2}$CHO$_{upper}$/CH$_{3}$OH  \\
            & cm$^{-2}$ & & & & \\
            \noalign{\smallskip}
            \hline
            \noalign{\smallskip}

         	Model F	& & 1.1	& 3.4E-02	& 3.1E-02 & 3.1E-02 \\
         	Model M	& & 1.7	& 1.0E-01	& 7.3E-02 & 7.3E-02 \\
         	Model S	& & 1.3	& 1.3E-01	& 1.0E-01 & 1.0E-01 \\
         	            \hline
            \noalign{\smallskip}
            AFGL 2591	& <1.9E+16 & -	& -  & <3.9E-01	& - \\
            G24.78		& <2.4E+16 & <3.3E+01 & <8.5E-02	& 2.6E-03 & 9.0E-04 \\
           	G31.41+0.31	& <5.8E+16 & <2.8E+01 & <4.9E-02	& 1.8E-03 & 3.8E-03 \\
           	G75.78		& <3.5E+16 & <1.7E+02 & <3.1E-01	& 1.8E-03 & 2.6E-02\\
           	IRAS 18089-1732	& <4.2E+16 & <5.0E+01 & <1.9E-01	& 3.8E-03 & 7.9E-03 \\
           	IRAS 20216+4104	& <6.4E+16 & -	& <2.2	& -  & -\\
           	NGC 7538 IRS1	& <2.0E+16 & <3.5E+01 & <1.8E-01	& 4.8E-03 &2.1E-04 \\
           	W3(H$_{2}$O)	& <5.0E+16 & <3.9E+01 & <5.0E-02	& 1.3E-03 &6.4E-04 \\
           	W3(H$_{2}$O)*	& <1.7E+16 & <1.3E+01 & <1.7E-02   & 1.3E-03 & 6.4E-04 \\
           	W 33A		& <5.7E+16 & <2.7E+01 & <2.9E-01	& 1.1E-02 & 4.6E-03\\
           	         	            \hline
            \noalign{\smallskip}
           	Sgr B2$^{a}$ & 1.2E+14 & 5.7E-01 & 1.7E-02 & 1.3E-02 \\
           	Sgr B2(M)$^{b}$ & 4.5E+16 & 3.2 & 1.7E-02 & 5.2E-03 \\
           	Sgr B2(N)$^{b}$ & 6.0E+17 & 4.3E-01 & 3.3E-02 & 7.8E-02 \\
           	Sgr B2(N)$^{c}$ & 5.0E+17 & 2.1 & 1.0E-01 & 4.8E-02 \\
        	Orion Compact Ridge$^{d}$		& -	& -	& - & 1.6E-03  \\ 	
          	           
            \noalign{\smallskip}
            \hline
            
         \end{tabular}     
         $$     

         \tablefoot{Column densities for the assumed source size and 
upper limit abundance ratios for methylamine, derived from the $2_2-2_1$ line assuming T$_{\rm rot}$ = 120 K. The values for NH$_{2}$CHO and CH$_{3}$OH where taken from \cite{bisschop2007} and \cite{isokoski2013}. The upper limits of NH$_{2}$CHO were determined by our own re-analysis of the \cite{bisschop2007} data and taken from the appendix of \cite{isokoski2013}.
 
\textbf{*} Column density calculated for the $8_2-8_1$ line.
         
        \textbf{References.} $^{a}$ \citet{turner1991}, beam sizes between 65" and 107", assuming no beam dilution; $^{b}$ \citet{belloche2013}, beam sizes between ~25" and ~30", assuming a $3''$ source size for (N) and $5''$ source size for (M);  $^{c}$ \citet{neill2014}, beam sizes between ~10" and ~40", assuming source size of 2.5$''$; and $^{d}$\citet{crockett2014} beam sizes between 44" and 11"  and assuming a $10''$ size of the Compact Ridge.}

   \end{table*}

      \begin{table}
      \caption[]{Upper limit column densities and abundance ratios for methylamine.}
         \label{ratioscya}
         $$
         \begin{tabular}{l l l}
            \hline
            \noalign{\smallskip}
            Source & CH$_{3}$NH$_{2}$/CH$_{3}$CN & CH$_{3}$CN/CH$_{3}$OH   \\
            
            \noalign{\smallskip}
            \hline
            \noalign{\smallskip}

         	Model F	&	5.5E+01 & 6.3E-04 \\
         	Model M	&	3.8E+01	& 2.6E-03 \\
         	Model S	&	1.5E+01 & 8.6E-03 \\
         	            \hline
            \noalign{\smallskip}
            AFGL 2591	& <5.3 & <7.5E-02\\
            G24.78		& <5.1E-01 & 2.1E-01\\
           	G31.41+0.31	& <3.6 & 5.9E-02*\\
           	G75.78		& <1.9E+01 & 1.6E-02\\
           	IRAS 18089-1732	& <8.9 & 1.1E-02*\\
           	IRAS 20216+4104	& <4.3E+01 & 5.2E-02\\
           	NGC 7538 IRS1	& <2.5 & 6.8E-02\\
           	W3(H$_{2}$O)	& <7.2 & 7.0E-03\\
           	W 33A		& <2.1 & 1.4E-01\\
           	         	            \hline
            \noalign{\smallskip}
           	Sgr B2$^{a}$ & 1.2 & 1.5E-02\\
           	Sgr B2(N)$^{b}$ & 3.0E-01 & 1.1E-01 \\
           	Sgr B2(M)$^{b}$ & 2.5E-01 & 6.7E-02 \\
           	Sgr B2(N)$^{c}$ & 5.9E+01 & 1.7E-02 \\
           	Orion Compact Ridge$^{d}$ & - & 1.1E-02 \\
           	    
            \noalign{\smallskip}
            \hline
            
         \end{tabular}     
         $$     
         \tablefoot{Abundance ratios for methylamine. The values for CH$_{3}$OH and CH$_{3}$CN were taken from \cite{bisschop2007} and \cite{isokoski2013}. \\
         *Ratio derived from optically thin $^{13}$C isotope.
         
        \textbf{References.} $^{a}$ \citet{turner1991}, $^{b}$ \citet{belloche2013}, $^{c}$ \citet{neill2014} and $^{d}$ \citet{crockett2014}.}
           
   \end{table}

\section{Discussion}
\label{discus}

Despite a significant number of succesfully identified molecules (see
Table \ref{fulltrans} in the Appendix for examples of W3(H$_{2}$O) and
G31.41+0.31), only upper limits were found for methylamine in the
various hot cores, limiting the conclusions that can be
drawn. Nevertheless, trends are seen in our abundance ratios. The
results suggest that theoretically predicted abundances for
both methylamine and formamide are too high. In contrast, acetonitrile
is found to be underproduced in the models. In the following, each of
these species is discussed individually.

\subsection{CH$_3$NH$_2$}

\citet{garrod2008} suggest that methylamine is primarily formed by grain-surface chemistry
using UV to create the CH$_3$ and NH$_2$ radicals from
photodissociation of primarily CH$_4$ and NH$_3$. Perhaps the amount
of UV processing is overestimated in these models. An alternative
route is hydrogen atom addition to solid HCN, proposed by
\citet{theule2011} and found to lead to both CH$_{2}$NH (methanimine)
and CH$_3$NH$_2$.  \citet{walsh2014} find in their models that
methylamine is indeed efficiently formed on grains at 10 K by atom
addition reactions to solid CH$_{2}$NH.  \citet{burgdorf2010} have
detected HCN ice on Triton, but so far no detection of solid HCN has
been made in the ISM.  Methanimine is actually readily observed in the
gas-phase \citep{turner1991,nummelin1998,belloche2013} so the presence
of both species makes the H-atom addition scheme probable. However,
\citet{halfen2013} detect CH$_2$NH in Sgr B2(N) at a rotational
temperature of 44 K, which is distinctly colder than the 159 K
observed for CH$_3$NH$_2$, suggesting that the two molecules may not
co-exist.  An alternative route would therefore be to form these
molecules by two different gas-phase reaction pathways
(CH$^{\bullet}$(g) + NH$_{3}$(g) $\rightarrow$ CH$_{2}$NH + H and
CH$_{3}^{\bullet}$(g) + NH$_{3}$(g) $\rightarrow$ CH$_{3}$NH$_{2}$ +
H), with CH being present primarily in the colder outer envelope and
CH$_3$ in the warmer center. Further modeling is needed to determine
whether these gas-phase reactions can reproduce the observed
abundances quantitatively.

\subsection{NH$_2$CHO}

Formamide also appears to be overproduced in the hot core model. Since
\citet{garrod2008} use both gas-phase, radical and atom addition
reactions to form formamide, it is difficult to pin down where the
discrepancies could come from. It is known that NH$_{2}$CHO is formed
in CO:NH$_{3}$ mixtures after UV and electron irradiation
\citep{grim1989, demyk1998, jones2011} and it has also been proposed
that it can form from H- and N-atom addition to solid CO
\citep{tielenscharnley1997}.  Gas-phase formation from CO and NH$_3$
is viable as well \citep{hubbard1975}, although these experiments were
conducted under high-pressure conditions, not the low pressures
applicable in the ISM.  Further quantification of both gas-phase and
solid phase routes through laboratory experiments is needed.  Recent
laboratory experiments by \citet{fedoseev2014} do not find NH$_2$CHO
production in H- and N-atom bombardment studies of solid CO,
consistent with a large barrier for H- addition to HNCO found in ab
initio calculations \citep{nguyen1996}, so perhaps the efficiency of
this route has been overestimated in the models.  An alternative
solution would be that the high-mass sources studied here have not
gone through a long (pre-stellar) phase in which the dust temperature
was low enough for CO to be frozen out and turned into other
molecules.

\subsection{CH$_3$CN}

The clear mismatches between theory and observations for the ratios
involving CH$_3$CN point toward an underproduction of acetonitrile by
more than an order of magnitude in the models. As with formamide,
gas-phase, radical and atom addition reactions contribute to the
formation of CH$_{3}$CN in the models, making it difficult to
determine the cause. The main formation route in the models
by radical addition of solid CH$_{3}^{\bullet}$ and solid
CN$^{\bullet}$ has never experimentally been investigated. It would
therefore be useful to determine if this is a viable solid state
formation route and if it potentially has a higher efficiency than
assumed.

Alternatively it is possible that photodestruction of solid
acetonitrile is not as efficient as assumed in the models. \citet{gratier2013} find high gas-phase CH$_{3}$CN abundances in the Horsehead PDR, indicative of a high photodesorption rate and slow destruction of CH$_{3}$CN in the ice. \citet{bernstein2004} indeed find slower photolysis of solid CH$_{3}$CN compared with other organic molecules. If such a slower photodissociation rate would also hold  for gas-phase CH$_3$CN, it would be an attractive explanation why the CH$_3$CN rotational temperatures are generally higher than those of other complex molecules (e.g., \citet{bisschop2007} and many other hot core studies), since the molecule could then approach the protostar closer before being destroyed. However, the gas phase photoabsorption cross sections of
CH$_3$CN are well determined and if the bulk of these absorptions lead
to dissociation this would result in a photodissociation rate of
gaseous CH$_3$CN at least as fast as that of CH$_3$OH
\citep{vandishoeck2006a}.

Another important parameter for all molecules studied here is the
mobility of radicals and neutral molecules on the surface assumed in
the gas-grain models. For many species no experimental data are
available on diffusion barriers, only theoretically-inspired
guesses. Observational evidence suggests that at least parts of the
ice mantles are segregated in CO-rich and CO-poor layers
\citep{tielens1991,pontoppidan2008}. Therefore, more knowledge of the
structure of ice mantles and the mobility of radicals and neutral
molecules as a function of surface temperature and in various chemical
environments is necessary to determine if addition reactions are
likely to happen and at which rates.

\subsection{Prospects for ALMA}

In the near future, much deeper searches for CH$_3$NH$_2$ can be
carried out by ALMA (Appendix \ref{app_alma}). Figure \ref{almaspec1} shows that the strongest transitions within Band 6 are mainly located between 240 and 275 GHz and in Band 7 around 310, 340 and 355 GHz. In Table
\ref{almameth} the strongest transitions in ALMA's Band 6 and 7 are listed. It becomes apparent that lines covered by Band 7 are more intense, but at the cost of a lower line density. 

Estimates done for the W 33A source with the CASSIS line analysis software and the ALMA Sensitivity Calculator show that ALMA should be able to reach the 3$\sigma$ detection limits for the CH$_3$NH$_2$ lines around 236 GHz in less than 1 hour of integration time, assuming the column density for methylamine of 1.2$ \times 10^{14}$ cm$^{-2}$ as found by \citet{turner1991} in a large beam and two orders of magnitude lower than those inferred here for a small source size. This estimate assumes a spectral resolution of 0.64 kms$^{-1}$ as used in our JCMT data, the number of ALMA antennas set to 34 (as in Cycle 2) and a synthesized beam of 1.1", appropriate for the W33A hot core (100 K radius).

\section{Conclusions}
\label{con}

We have analysed nine hot core regions in search of methylamine.  The
molecule has not been convincingly detected, so upper limit abundances
are determined for all the sources. From these limits, ratios of
methylamine to other molecules (NH$_2$CHO, CH$_3$OH, CH$_3$CN) have
been determined and compared with theory and Sagittarius B2
surveys. Our conclusions are as follows:

   \begin{enumerate}
   
      \item Trends in our results indicate that both
        methylamine and formamide are overproduced in the models of
        \citet{garrod2008}. Acetonitrile is underproduced with respect
        to these models. This is especially true for the slow models.

        \item Abundance ratios do not differ more than an order of magnitude between various sources suggesting that the
          (nitrogen) chemistry is very similar between hot cores, as
          has been found previously for other species.

      \item More (laboratory) studies are needed to clarify the formation
        pathway of methylamine and to determine differences and
        similarities with formamide, methanimine and, to a lesser
        extent, acetonitrile formation.

      \item The upper limits determined for CH$_{3}$NH$_{2}$ here can
        guide future more sensitive observations, especially with
        ALMA. Based on the ratios found in the Sgr B2 observations it
        is very likely that ALMA will reach the detection limit for
        methylamine in the sources studied here. Particularly strong transitions and
  spectral regions to target with ALMA are given.
  
 \end{enumerate}

\begin{acknowledgements}
  We would like to thank C. Walsh, I. San Jose Garc\'{i}a,
  M. Drozdovskaya, N. van der Marel, M. Kama, M. Persson, J. Mottram,
  G. Fedoseev, and H. Linnartz for their support and input on this
  project.  Thoughtful comments by the referee are much
  appreciated. Astrochemistry in Leiden is supported by the
  Netherlands Research School for Astronomy (NOVA), by a Royal
  Netherlands Academy of Arts and Sciences (KNAW) professor prize, and
  by the European Union A-ERC grant 291141 CHEMPLAN.
\end{acknowledgements}


\bibliographystyle{aa}
\bibliography{lib}

\newpage

\Online

\begin{appendix}
\section{Methylamine and other transitions}
\label{app_trans}

   \begin{figure*}
   \centering
   \includegraphics[width=\hsize]{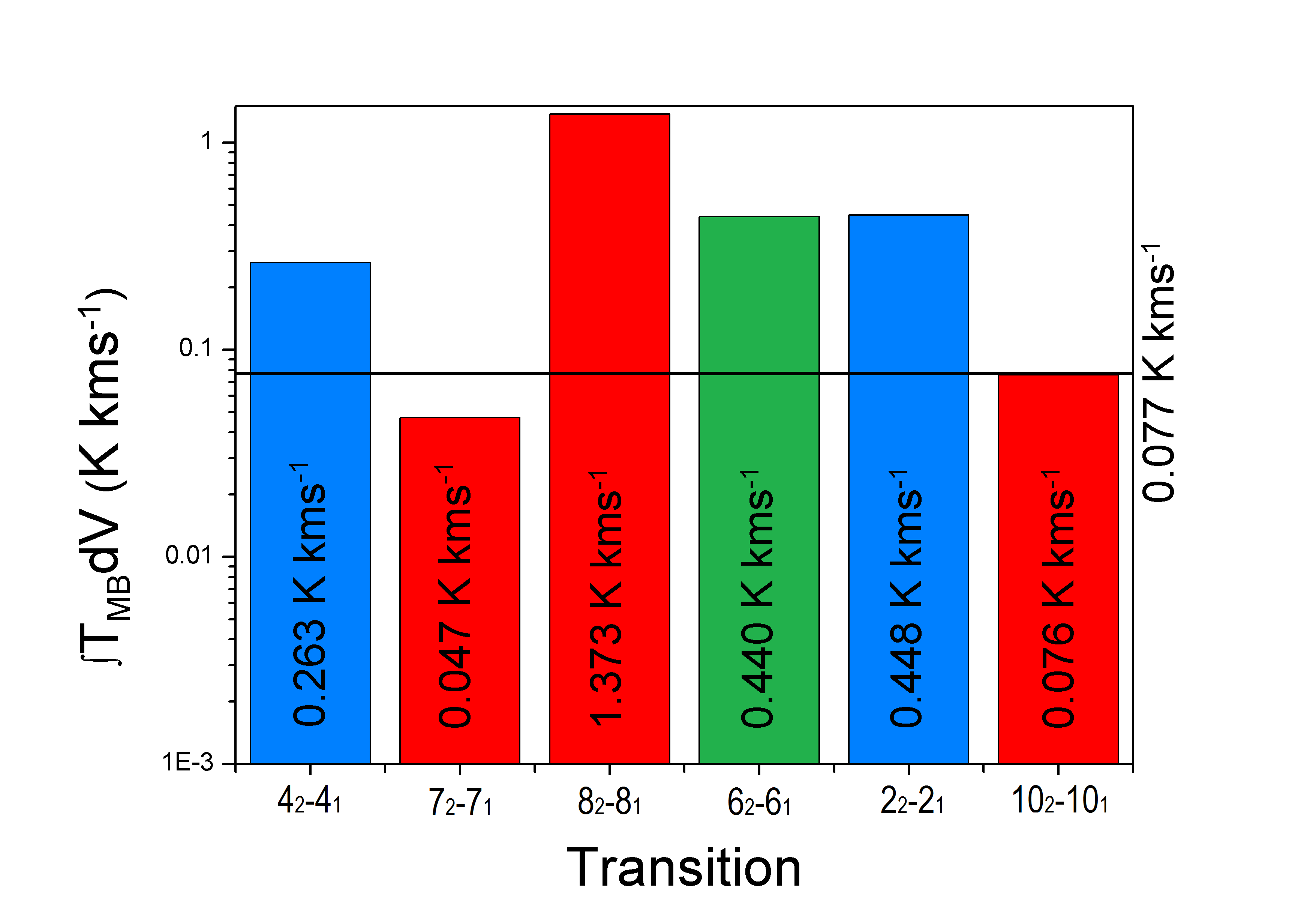}
      \caption{Bargraph plot of the integrated main-beam intensities
        for the six investigated methylamine
        transitions. \textit{$\int T_{ \rm{MB}}dV$} values were
        calculated for a total column density of 3.4 $\times$
        10$^{17}$ cm$^{-2}$ as inferred toward G31.41+0.31 from the
        6$_{2}$ $\rightarrow$ 6$_{1}$ transition, assuming a
        rotational temperature of 120 K. The horizontal line shows
        the 0.077 K kms$^{-1}$ 3$\sigma$ value for G31.41+0.31. It is
        unlikely that the detection of the 236408 feature (green) in
        this hot core is methylamine, since the main-beam
        intensities of other CH$_{3}$NH$_{2}$ transitions are above the
        3$\sigma$ value. Particularly the 4$_{2}$ $\rightarrow$
        4$_{1}$ and 2$_{2}$ $\rightarrow$ 2$_{1}$ transition (blue)
        should be visible in our spectra. The remaining transitions
        (red) are either below detection limit or not observed toward
        this source.  }
         \label{tmb}
   \end{figure*}

   \begin{table*}
      \caption[]{All identified transitions for the sources G31.41+0.31 and W3(H$_{2}$O), with integrated peak area listed.}
         \label{fulltrans}
     $$ 
         \begin{tabular}{l l r r c r r}
            \hline
            \noalign{\smallskip}
            Species & Freq	& $E_{\rm{up}}$	& \textbf{$A$} & Transition & G31.41+0.31 & W3(H$_{2}$O)  \\
            & (MHz)	& (K)	& (s$^{-1}$) & & (K kms$^{-1}$) & (K kms$^{-1}$)		\\
            \noalign{\smallskip}
            \hline
            \noalign{\smallskip}
            
            CH$_{3}$OCHO & 228628.876 & 118.8 & 1.66E-04 & 18$_{513 \: 2}$ $\rightarrow$ 17$_{512 \: 2}$ & - & 0.39     \\
            CH$_{3}$OCHO & 228651.404 & 118.8 & 1.66E-04 & 18$_{513 \: 0}$ $\rightarrow$ 17$_{512 \: 0}$ & - & 0.53     \\
            CH$_{3}$COCH$_{3}$ & 228668.358 & 85.4 & 1.69E-04 & 14$_{9 \: 6 \: 1}$ $\rightarrow$ 13$_{8 \: 5 \: 2}$ & - & 0.24     \\
			CH$_{3}$OCHO & 229388.947 & 217.0 & 1.62E-05 & 23$_{915 \: 0}$ $\rightarrow$ 23$_{816 \: 0}$ & 0.79 & - \\            
			CH$_{3}$OCHO & 229405.021 & 110.7 & 1.75E-04 & 18$_{315 \: 2}$ $\rightarrow$ 17$_{314 \: 2}$ & 2.18 & 0.78 \\            
			CH$_{3}$OCHO & 229420.342 & 110.7 & 1.75E-04 & 18$_{315 \: 0}$ $\rightarrow$ 23$_{314 \: 0}$ & 1.86 & 0.71 \\            
			CH$_{3}$OCHO & 229504.724 & 134.3 & 1.18E-05 & 20$_{317 \: 0}$ $\rightarrow$ 19$_{416 \: 0}$ & 1.52 & - \\            
			CH$_{3}$OH & 229589.056 & 374.4 & 2.08E-05 & 15$_{4 \: 0}$ $\rightarrow$ 16$_{3 \: 0}$ & 2.58 & 1.06 \\            
			CH$_{3}$OH & 229758.756 & 89.1 & 4.19E-05 & 8$_{1 \: 0}$ $\rightarrow$ 7$_{0 \: 0}$ & 7.17 & - \\     
			CH$_{3}$OCHO & 236355.948 & 128.0 & 1.93E-04 & 20$_{318 \: 1}$ $\rightarrow$ 19$_{317 \: 1}$ & 2.38 & - \\                  
			CH$_{3}$OCHO & 236365.574 & 128.0 & 1.93E-04 & 20$_{318 \: 0}$ $\rightarrow$ 19$_{317 \: 0}$ & 1.93 & - \\            
			CH$_{3}$NH$_{2}$ & 236408.788 & 60.8 & 5.94E-05 & 6$_{2 \: 1}$ $\rightarrow$ 6$_{1 \: 0}$ & 0.44 & - \\
			HCCCN & 236512.777 & 153.2 & 1.05E-03 & 26 $\rightarrow$ 25 & 4.67 & 2.27 \\
			H$_{2}$CS & 236726.770 & 58.6 & 1.91E-04 & 7$_{1 \: 7}$ $\rightarrow$ 6$_{1 \: 6}$ & 7.54 & 5.75 \\
			CH$_{3}$OCHO & 236743.697 & 129.6 & 1.86E-04 & 19$_{515 \: 1}$ $\rightarrow$ 18$_{514 \: 1}$ & 2.21 & ?$^{a}$ \\            
			CH$_{3}$OCHO & 236759.687 & 129.6 & 1.86E-04 & 19$_{515 \: 0}$ $\rightarrow$ 18$_{514 \: 0}$ & 1.40 & 0.53 \\            
			CH$_{3}$OCHO & 236800.589 & 136.7 & 1.80E-04 & 19$_{614\: 1}$ $\rightarrow$ 18$_{613 \: 1}$ & 1.21 & ? \\            
			CH$_{3}$OCHO & 236810.314 & 136.7 & 1.81E-04 & 19$_{614 \: 0}$ $\rightarrow$ 18$_{613 \: 0}$ & 2.50 & 1.20 \\            
			CH$_{3}$OH & 236936.089 & 260.2 & 2.79E-05 & 14$_{1 \: 0}$ $\rightarrow$ 13$_{2 \: 0}$ & 2.74 & 1.91 \\            
			CH$_{3}$OCHO & 236975.844 & 320.3 & 2.01E-04 & 22$_{122 \: 4}$ $\rightarrow$ 21$_{121 \: 4}$ & 0.86 & 0.33 \\            
			CH$_{3}$OCHO & 236976.390 & 320.3 & 2.01E-04 & 22$_{022 \: 5}$ $\rightarrow$ 21$_{021 \: 0}$ & 0.86 & 0.33 \\            
			CH$_{3}$OCHO & 236975.844 & 320.3 & 2.01E-04 & 22$_{122 \: 4}$ $\rightarrow$ 21$_{121 \: 4}$ & 1.19 & 0.52 \\            
            CH$_{3}$OCHO & 236976.390 & 320.3 & 2.01E-04 & 22$_{022 \: 5}$ $\rightarrow$ 21$_{021 \: 5}$ & 1.19 & 0.52 \\            
            CH$_{3}$OCH$_{3}$ & 237046.092 & 31.3 & 2.33E-05 & 7$_{2 \: 5 \: 3}$ $\rightarrow$ 6$_{1 \: 6 \: 3}$ & 2.90 & 0.80 \\    
            CH$_{3}$OCH$_{3}$ & 237046.106 & 31.3 & 2.33E-05 & 7$_{2 \: 5 \: 2}$ $\rightarrow$ 6$_{1 \: 6 \: 2}$ & 2.90 & 0.80 \\       
            CH$_{3}$OCH$_{3}$ & 237048.797 & 31.3 & 2.32E-05 & 7$_{2 \: 5 \: 1}$ $\rightarrow$ 6$_{1 \: 6 \: 1}$ & 2.90 & 0.80 \\    
            CH$_{3}$OCH$_{3}$ & 237051.495 & 31.3 & 2.33E-05 & 7$_{2 \: 5 \: 0}$ $\rightarrow$ 6$_{1 \: 6 \: 0}$ & 2.90 & 0.80 \\  
            SO$_{2}$ & 237068.870 & 94.0 & 1.14E-04 & 12$_{3 \: 9}$ $\rightarrow$ 12$_{2 \: 10}$ & 1.07 & 1.96 \\  
            OC$^{34}$S & 237273.635 & 119.6 & 3.88E-05 & 20$\rightarrow$ 19 & 1.24 & 0.58 \\ 
            CH$_{3}$OCHO & 237297.482 & 128.0 & 1.95E-04 & 20$_{218 \: 2}$ $\rightarrow$ 19$_{217 \: 2}$ & ? & 2.45 \\ 
            CH$_{3}$OCHO & 237309.540 & 131.6 & 1.98E-04 & 21$_{220 \: 1}$ $\rightarrow$ 20$_{219 \: 1}$ & ? & 2.45 \\ 
            CH$_{3}$OCHO & 237315.082 & 131.6 & 1.98E-04 & 21$_{220 \: 0}$ $\rightarrow$ 20$_{219 \: 0}$ & ? & 2.45 \\   
            CH$_{3}$OCHO & 237344.870 & 131.6 & 1.98E-04 & 21$_{120 \: 2}$ $\rightarrow$ 20$_{119 \: 2}$ & 3.48 & 1.19 \\ 
            CH$_{3}$OCHO & 237350.386 & 131.6 & 1.98E-04 & 21$_{120 \: 0}$ $\rightarrow$ 20$_{119 \: 0}$ & 3.48 & 1.19 \\ 
            OCS & 243218.040 & 122.6 & 4.18E-05 & 20 $\rightarrow$ 19 & - & 2.95 \\ 
            CH$_{3}$OCH$_{3}$ & 259982.561 & 226.6 & 7.27E-05 & 20$_{516 \: 2}$ $\rightarrow$ 20$_{417 \: 2}$ & - & 1.54 \\ 
            CH$_{3}$OCH$_{3}$ & 259982.596 & 226.6 & 7.27E-05 & 20$_{516 \: 3}$ $\rightarrow$ 20$_{417 \: 3}$ & - & 1.54 \\ 
            CH$_{3}$OCH$_{3}$ & 259984.480 & 226.6 & 7.27E-05 & 20$_{516 \: 1}$ $\rightarrow$ 20$_{417 \: 1}$ & - & 1.54 \\ 
            CH$_{3}$OCH$_{3}$ & 259982.561 & 226.6 & 7.27E-05 & 20$_{516 \: 0}$ $\rightarrow$ 20$_{417 \: 0}$ & - & 1.54 \\ 
            NH$_{2}$CHO & 260189.848 & 92.4 & 1.25E-03 & 12$_{210}$ $\rightarrow$ 11$_{2 \: 9}$ & - & 1.46 \\ 
            H$^{13}$CO+ & 260255.339 & 25.0 & 1.33E-03 & 3 $\rightarrow$ 2 & - & 11.76 \\ 
            CH$_{3}$OCH$_{3}$ & 260327.165 & 208.3 & 7.21E-05 & 19$_{515 \: 2}$ $\rightarrow$ 19$_{416 \: 2}$ & - & 1.27 \\
            CH$_{3}$OCH$_{3}$ & 260327.238 & 208.3 & 7.21E-05 & 19$_{515 \: 3}$ $\rightarrow$ 19$_{416 \: 3}$ & - & 1.27 \\
            CH$_{3}$OCH$_{3}$ & 260329.312 & 208.3 & 7.21E-05 & 19$_{515 \: 1}$ $\rightarrow$ 19$_{416 \: 1}$ & - & 1.27 \\
            CH$_{3}$OCH$_{3}$ & 260331.422 & 208.3 & 7.21E-05 & 19$_{515 \: 0}$ $\rightarrow$ 19$_{416 \: 0}$ & - & 1.27 \\
            SiO & 260518.020 & 43.8 & 7.21E-05 & 6 $\rightarrow$ 5 & - & 9.24 \\
                     
            \noalign{\smallskip}
            \hline
         \end{tabular}
     $$ 
     $^{a}$ Indicates an identified transition where it was not possible to determine the peak area.
   \end{table*}
   
      \begin{figure*}
   \centering
   \includegraphics[width=\hsize]{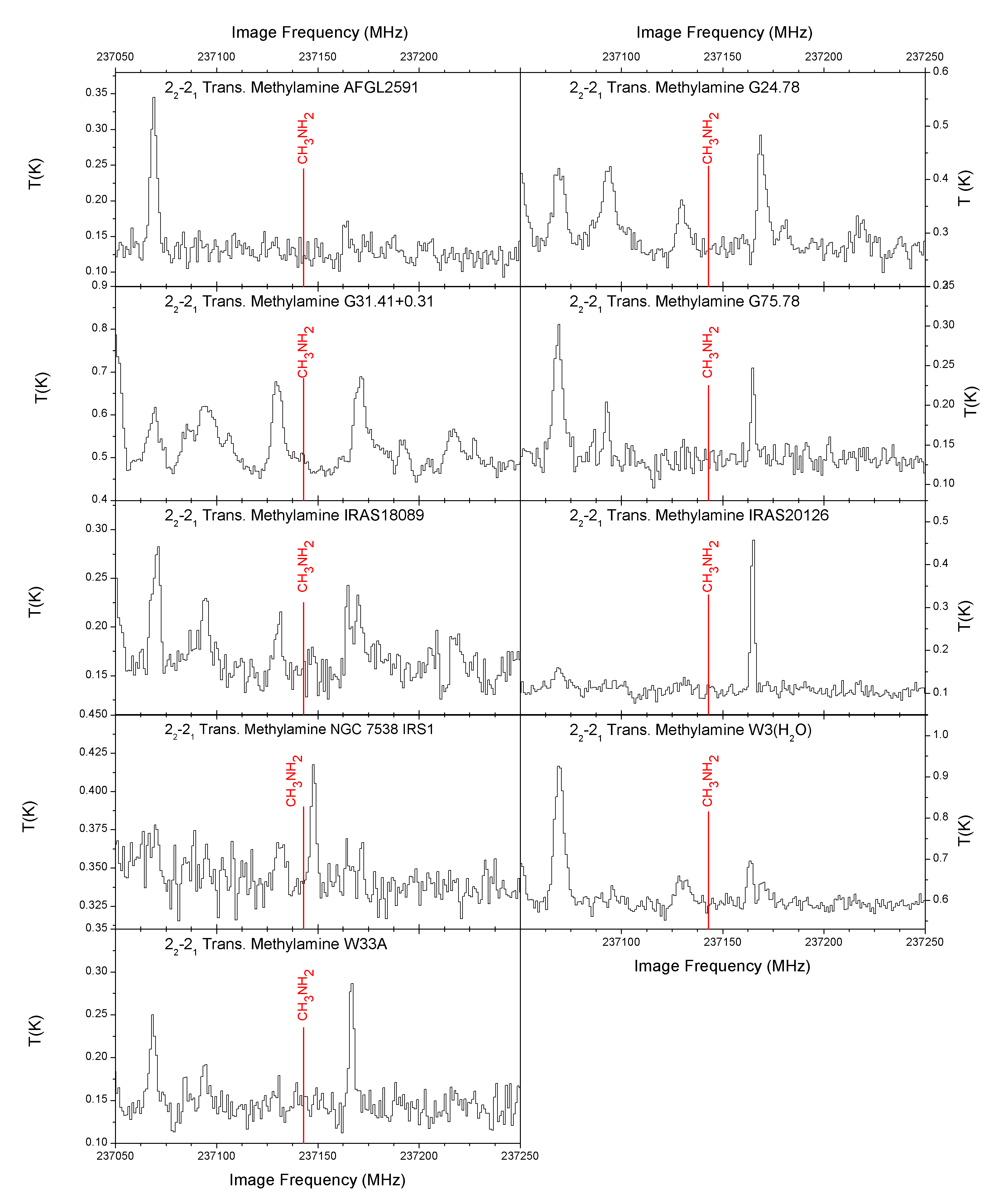}
      \caption{Blow-up of the spectral region around the 2$_{2}$ $\rightarrow$ 2$_{1}$ transition at 237143 MHz of all analysed hot cores. Despite being a particularly strong transition, it was not observed in any of the spectra. }
         \label{2-2}
   \end{figure*}

\onecolumn
\section{ALMA}
\label{app_alma}

In the following figure the simulated spectrum of methylamine is shown for a column density of 1.0 $\times$ 10$^{15}$ cm$^{-2}$ at an excitation temperature of 120 K. Simulations were done with CASSIS using the JPL spectroscopic database. The frequency ranges were taken to cover ALMA Bands 6 and 7. The resolution was set at 0.1 MHz for this spectrum.

      \begin{figure}[!h]
		\centering
    \includegraphics[width=\hsize]{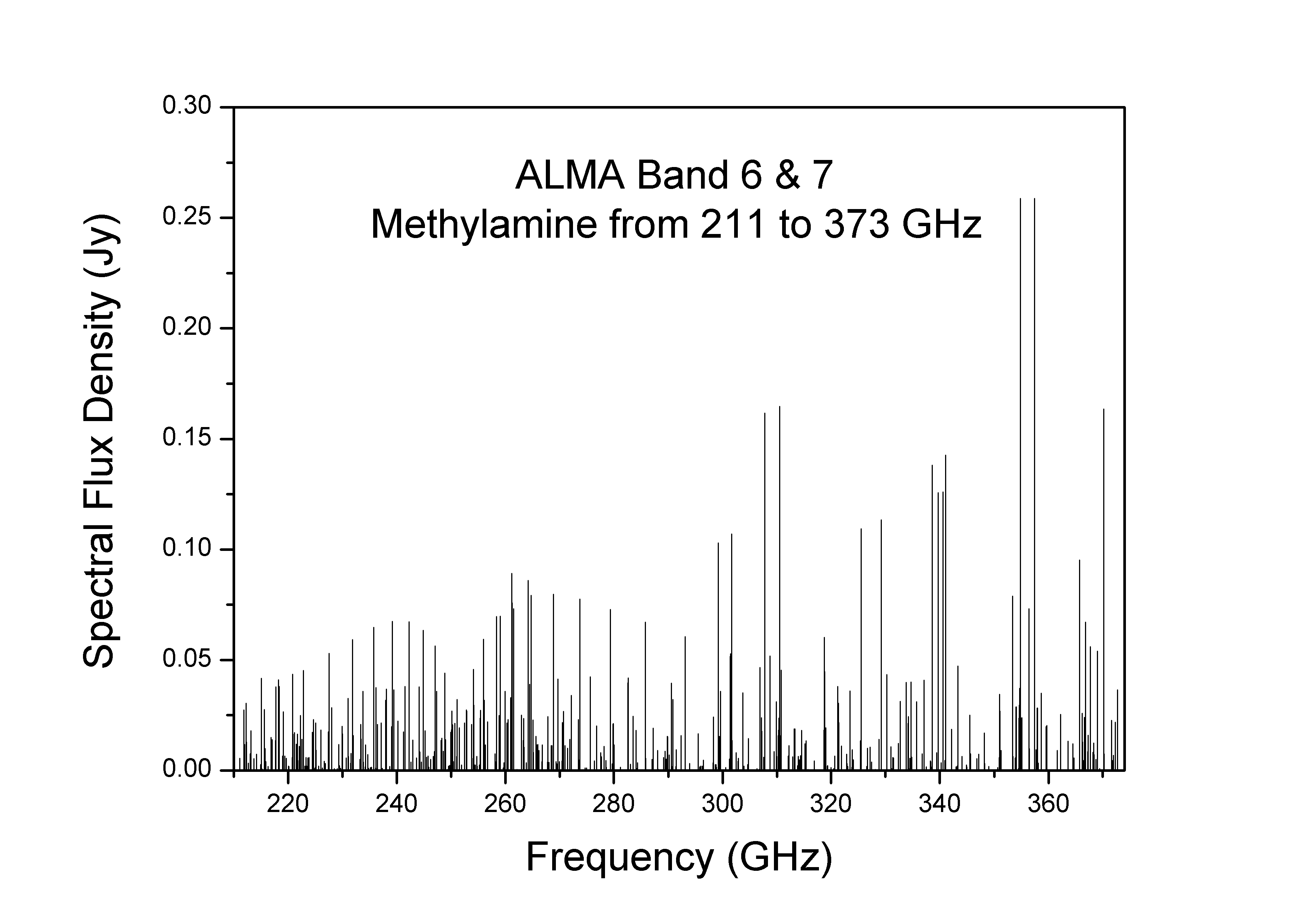}
      \caption{211 to 373 GHz spectrum of methylamine, covering ALMA bands 6 and 7. A column density of 10$^{15}$ cm$^{-2}$, T$_{\rm rot}$ and beam size of 1" are used.}
     \label{almaspec1}
   \end{figure}

      \begin{table}
      \caption[]{Methylamine transition target candidates for ALMA Band 6 and 7}
         \label{almameth}
     $$ 
         \begin{tabular}{c c r c}
            \hline
            \noalign{\smallskip}
            Transition &  Freq	& $E_{\rm{up}}$	& \textbf{$A$} \\
            & (MHz)	& (K)	& (s$^{-1}$) \\
            \noalign{\smallskip}
            \hline
            \noalign{\smallskip}
            
            7$_{2\ 6}$ $\rightarrow$ 7$_{1\ 6}$ &  227498.06 & 75.4 & 4.80E-05 \\
            5$_{2\ 6}$ $\rightarrow$ 5$_{1\ 6}$ &  232003.95 & 47.7 & 4.73E-05 \\
            8$_{2\ 3}$ $\rightarrow$ 8$_{1\ 2}$ &  235735.04 & 92.8 & 6.13E-05 \\
            7$_{2\ 2}$ $\rightarrow$ 7$_{1\ 3}$ &  239209.63 & 75.8 & 6.29E-05 \\
            6$_{2\ 3}$ $\rightarrow$ 6$_{1\ 2}$ &  242262.02 & 60.9 & 6.39E-05 \\
            5$_{2\ 2}$ $\rightarrow$ 5$_{1\ 3}$ &  244886.90 & 48.1 & 6.42E-05 \\
            8$_{0\ 5}$ $\rightarrow$ 7$_{1\ 5}$ &  259042.46 & 77.0 & 5.73E-05 \\            
            6$_{2\ 2}$ $\rightarrow$ 6$_{1\ 3}$ &  261252.89 & 60.9 & 7.14E-05 \\
            8$_{0\ 2}$ $\rightarrow$ 7$_{1\ 3}$ &  261563.15 & 76.8 & 5.98E-05 \\
            4$_{1\ 5}$ $\rightarrow$ 3$_{0\ 5}$ &  264172.21 & 25.9 & 8.74E-05 \\
            8$_{2\ 2}$ $\rightarrow$ 8$_{1\ 3}$ &  268898.14 & 92.8 & 7.44E-05 \\        
            9$_{0\ 5}$ $\rightarrow$ 8$_{1\ 5}$ &  299189.80 & 96.1 & 8.78E-05 \\
            9$_{0\ 3}$ $\rightarrow$ 8$_{1\ 2}$ &  301654091 & 95.9 & 9.11E-05 \\
            5$_{-1\ 3}$ $\rightarrow$ 4$_{0\ 2}$ &  307791.75 & 36.3 & 1.44E-04 \\
            14$_{2\ 2}$ $\rightarrow$ 14$_{-1\ 3}$ &  310750.84 & 240.0 & 8.33E-05 \\
            10$_{0\ 5}$ $\rightarrow$ 9$_{1\ 5}$ &  338628.32 & 117.3 & 1.25E-04 \\
            10$_{0\ 2}$ $\rightarrow$ 9$_{-1\ 3}$ &  341059.48 & 117.1 & 1.30E-04 \\
            6$_{1\ 2}$ $\rightarrow$ 5$_{0\ 3}$ &  354843.73 & 49.2 & 2.16E-04 \\
            6$_{1\ 5}$ $\rightarrow$ 5$_{0\ 5}$ &  357440.12 & 49.5 & 2.16E-04 \\
            3$_{2\ 4}$ $\rightarrow$ 2$_{1\ 4}$ &  370166.34 & 28.4 & 2.11E-04 \\

            \noalign{\smallskip}
            \hline
         \end{tabular}
     $$ 

   \end{table}

\end{appendix}
\end{document}